\documentstyle[12pt,oldlfont]{article}
\newcommand{\n}{\hspace*{-2.5mm}}
\newcommand{\simgt}{\rlap{\lower 3.5 pt \hbox{$\mathchar \sim$}} \raise 1pt
 \hbox {$>$}}
\newcommand{\simlt}{\rlap{\lower 3.5 pt \hbox{$\mathchar \sim$}} \raise 1pt
 \hbox {$<$}}

\catcode`@=11
\newcount\@tempcntc
\def\@citex[#1]#2{\if@filesw\immediate\write\@auxout{\string\citation{#2}}\fi
  \@tempcnta\z@\@tempcntb\m@ne\def\@citea{}\@cite{\@for\@citeb:=#2\do
    {\@ifundefined
       {b@\@citeb}{\@citeo\@tempcntb\m@ne\@citea\def\@citea{,}{\bf ?}\@warning
       {Citation `\@citeb' on page \thepage \space undefined}}%
    {\setbox\z@\hbox{\global\@tempcntc0\csname b@\@citeb\endcsname\relax}%
     \ifnum\@tempcntc=\z@ \@citeo\@tempcntb\m@ne
       \@citea\def\@citea{,}\hbox{\csname b@\@citeb\endcsname}%
     \else
      \advance\@tempcntb\@ne
      \ifnum\@tempcntb=\@tempcntc
      \else\advance\@tempcntb\m@ne\@citeo
      \@tempcnta\@tempcntc\@tempcntb\@tempcntc\fi\fi}}\@citeo}{#1}}
\def\@citeo{\ifnum\@tempcnta>\@tempcntb\else\@citea\def\@citea{,}%
  \ifnum\@tempcnta=\@tempcntb\the\@tempcnta\else
   {\advance\@tempcnta\@ne\ifnum\@tempcnta=\@tempcntb \else \def\@citea{--}\fi
    \advance\@tempcnta\m@ne\the\@tempcnta\@citea\the\@tempcntb}\fi\fi}
\catcode`@=12
\begin{document}
\title{\vskip-3cm{\baselineskip14pt
\centerline{\normalsize DESY 95--114\hfill ISSN 0418-9833}
\centerline{\normalsize FERMILAB--PUB--95/154--T\hfill}
\centerline{\normalsize MPI/PhT/95--54\hfill}
\centerline{\normalsize hep-ph/9506437\hfill}
\centerline{\normalsize June 1995\hfill}}
\vskip1.5cm
Neutral-Kaon Production in $e^+e^-$, $ep$, and $p\bar{p}$ Collisions at
Next-to-Leading Order}
\author{J. Binnewies$^1$, B.A. Kniehl$^2$\thanks{Permanent address:
Max-Planck-Institut f\"ur Physik, Werner-Heisenberg-Institut,
F\"ohringer Ring 6, 80805 Munich, Germany.}, and G. Kramer$^1$\\
$^1$ II. Institut f\"ur Theoretische Physik\thanks{Supported
by Bundesministerium f\"ur Forschung und Technologie, Bonn, Germany,
under Contract 05~6~HH~93P~(5),
and by EEC Program {\it Human Capital and Mobility} through Network
{\it Physics at High Energy Colliders} under Contract
CHRX-CT93-0357 (DG12 COMA).},
Universit\"at Hamburg\\
Luruper Chaussee 149, 22761 Hamburg, Germany\\
$^2$ Theoretical Physics Department, Fermi National Accelerator Laboratory\\
P.O. Box 500, Batavia, IL 60510, USA}
\date{}
\maketitle
\begin{abstract}
We present new sets of fragmentation functions for neutral kaons, both at
leading and next-to-leading order. They are fitted to data on inclusive
$K^0$~production in $e^+e^-$~annihilation taken by MARK~II at PEP
($\sqrt{s}=29$~GeV) and by ALEPH at LEP. Our fragmentation
functions lead to a good description of other $e^+e^-$~data on inclusive
$K^0$~production at various energies.
They also nicely agree with the $K_S^0$ transverse-momentum spectra measured
by H1 at the DESY $ep$~collider HERA, by UA5 at the CERN S$p\bar{p}$S Collider,
and by CDF at the Fermilab Tevatron.
\end{abstract}
\newpage

\section{Introduction}

Recently, precise data on inclusive $\pi^\pm$, $K^\pm$, and
unspecified-charged-hadron production in $e^+e^-$ annihilation at the
$Z$~resonance has been published.
Using this new data and similar data from lower centre-of-mass (CM) energy
($\sqrt{s}=29$~GeV), we constructed new
sets of fragmentation functions (FF) for charged pions and kaons at leading
order (LO) and next-to-leading order (NLO) \cite{bkk2}. These new
parameterizations were tested against data on $\pi^{\pm}$, $K^{\pm}$, and
charged-hadron production in $e^+e^-$ annihilation at various energies and
data on single-charged-hadron production in small-$Q^2$ $ep$ scattering
at HERA, which presents a
nontrivial check of the factorization theorem of the QCD-improved parton model.

Besides charged pions and kaons or just charged hadrons, $K_S^0$ mesons are
easily detected through their dominant decay into $\pi^+\pi^-$ pairs.
The ALEPH \cite{aleph}, DELPHI \cite{delphi}, OPAL \cite{opal}, and L3
\cite{l3}
Collaborations at LEP have recently reported on their high-statistics analyses
of inclusive single $K^0$ production.\footnote{Unless stated otherwise,
we shall collectively use the symbol $K^0$ for the sum of $K_S^0$ and $K_L^0$
(or $K^0$ and $\overline K^0$).}

Following our strategy of constructing FF for charged pions and kaons, we
shall combine this new data on $K^0$~production at the $Z$~resonance with the
rather precise data taken at $\sqrt s=29$~GeV by the MARK~II Collaboration
\cite{markii} at PEP to obtain FF for the neutral kaons.
Owing to the factorization theorem, the same FF can be used to
predict cross sections of inclusive single $K^0$ production at high transverse
momenta ($p_T$) in other processes like $ep$ and $p\bar{p}$ scattering.
The functions characterizing the fragmentation of gluons, $u$, $d$, $s$, $c$,
and $b$ quarks (antiquarks) into neutral kaons contribute quite differently
in these processes as compared to $e^+e^-$~annihilation. For example, in
$e^+e^-$~annihilation at the $Z$ resonance, all five quarks are directly
produced, whereas the gluon does not directly couple to the electroweak
currents. The gluon only contributes in higher orders and mixes with the
quarks through the $Q^2$~evolution. On the other hand,
in the case of inclusive light-meson production at moderate $p_T$ in
high-energy $p\bar{p}$~collisions, the cross section is
dominated by gluon fragmentation \cite{borzumati}. In $ep$ collisions with
almost real photons at HERA, the situation is mixed.
In the lower $p_T$ range ($p_T\simlt15$~GeV),
inclusive single hadron production proceeds dominantly via the resolved
photoproduction processes $gg\to gg$, $qg\to gq$, and $qg\to qg$, where the
first and second partons originate from the virtual photon and the proton,
respectively, while the third one fragments into the outgoing hadron
\cite{bkk}. Direct photoproduction only plays a significant r\^ole at larger
$p_T$ \cite{kk}. Therefore, quark and gluon fragmentation should give
comparable contributions even at small~$p_T$.

In our previous work on FF for charged pions and kaons \cite{bkk2},
we could exploit the information from tagged three-jet events in $e^+e^-$
annihilation to constrain the gluon fragmentation into charged hadrons, which
constrained also that into charged pions and kaons. Unfortunately, such
information is not yet available for inclusive $K^0$~production in
$e^+e^-$~annihilation. Thus, we shall have to resort to the information on
gluon fragmentation into charged kaons which we extracted in Ref.~\cite{bkk2}.

Another problem that requires special attention is related to the distinction
of the different quark flavours in $K^0$~fragmentation. In our recent analysis
of $\pi^\pm$ and $K^\pm$ fragmentation \cite{bkk2}, we
had some information on the fragmentation of specific flavours at our disposal.
Preliminary measurements of charged-hadron production by the ALEPH
Collaboration \cite{cowan}
distinguished between three cases, namely the fragmentation of (i)
$u$, $d$, $s$~quarks, (ii) $b$~quarks only, and (iii) all five flavours
($u$, $d$, $s$, $c$, and $b$). This enabled us to remove the assumption
that the $s$, $c$, and $b$~quarks fragment into charged pions (kaons) in the
same way, which we had made in our earlier work \cite{bkk1}.
Although equivalent information is still lacking for $K^0$~fragmentation
in $e^+e^-$~annihilation,
we shall follow the approach of our recent work on $\pi^\pm$ and $K^\pm$
production \cite{bkk2},
where no additional identities between the FF of different quark flavours were
imposed, except those following from the flavour content of the produced
mesons.
Should it turn out that the relative importance of the different flavours
cannot yet be pinned down so reliably, then this will not be due to a
shortcoming of this specific procedure;
this would just signal that more detailed data is indispensable in order to
determine the differences in flavour of the FF more accurately,
leaving room for further improvements.

It is the purpose of this work to make use of the new $K^0$ data by ALEPH
\cite{aleph} together with the $K^0$ data by MARK~II \cite{markii} to
construct new LO and NLO sets of FF, only identifying the FF of the $d$
and $s$~quarks and imposing no constraints on the quarks otherwise.
At the starting scale, $Q_0$, we shall take the gluon FF of the neutral kaons
to be equal to their charged counterparts.
The recent data from DELPHI \cite{delphi}, OPAL \cite{opal}, and
L3 \cite{l3} agree with the ALEPH data and will not be used in our
fit. A comparison of all four data sets may be found in a report by OPAL
\cite{opal}. We choose the ALEPH data, since, in the region of
relatively large $x$, which we are mainly interested in, it has a slightly
smaller total error than the data from DELPHI and OPAL.
The data from L3 does not extend to $x$ values in excess of 0.24 and is thus
less useful for our purposes.

Our new $K^0$ FF sets will be tested against older data from $e^+e^-$~colliders
with lower CM energies. Furthermore, we shall calculate the $p_T$ distributions
of $K_S^0$ mesons produced inclusively in $ep$ and $p\bar{p}$ collisions at
various CM energies and compare them with preliminary H1 data
\cite{linsel} and with data from UA5 \cite{ua5} and CDF \cite{cdf},
in order to check whether the gluon fragmentation and the relative importance
of the various quark flavours are realistically described.

An alternative way of constructing FF is to fit to data generated by
well-established Monte Carlo event generators rather than experimental data.
This avenue has just recently been taken in Ref.~\cite{gre}, where, among
other things, a NLO set of $K_S^0$ FF has been presented.
This offers us yet another opportunity to test our $K^0$ FF, namely against
Monte Carlo output.
We shall report the outcome of such a comparison later on.

The LO and NLO formalisms for extracting FF from $e^+e^-$~data are
comprehensively described in our previous works \cite{bkk2,bkk1} and will not
be reviewed here.
Also, the formulae that are needed to calculate the cross sections of inclusive
single hadron production in $ep$ collisions (with almost real photons) and in
$p\bar{p}$ collisions may be found in earlier publications
\cite{bkk2,bkk,kk}. The NLO formulae in these references are based on the
works by Aversa {\it et al.} \cite{aversa} (resolved photoproduction and
$p\bar{p}$~collisions), Aurenche {\it et al.} \cite{aurenche}
(direct photoproduction), and Altarelli {\it et al.} \cite{alt}
($e^+e^-$ collisions).

This paper is organized as follows.
In Sect.~2, we shall describe the actual analysis and present our results for
the $K^0$~FF. We shall also check these FF against $e^+e^-$~data at lower
energies which we did not use in our fits.
Furthermore, we shall compare the calculated
inclusive $K_S^0$ cross sections for $ep$ and $p\bar{p}$~collisions with
H1, UA5, and CDF experimental results. Our conclusions will be summarized in
Sect.~3. In the Appendix, we shall list simple parameterizations of our FF sets
for inclusive $K^0$~production.

\section{Results}

For our analysis, we select the data on $K^0$~production taken at energy
$\sqrt{s}=29$~GeV by the MARK~II Collaboration at PEP \cite{markii} and those
collected at $\sqrt{s}=M_Z$ by the ALEPH Collaboration at LEP \cite{aleph}.
This data comes in the form $(1/\sigma_{\rm had})d\sigma/dx$ as a
function of $x=2E_{K^0}/\sqrt{s}$, where $\sqrt{s}$ and $E_{K^0}$ are the
$e^+e^-$ and $K^0$ energies in the CM system, respectively.
The data from MARK~II and ALEPH lie within the ranges
$0.036\le x\le 0.69$ and $0.003698\le x\le 0.8187$, respectively.
These and other $e^+e^-$ experiments present inclusive cross sections for
$K_S^0+K_L^0$ (or $K^0+\overline{K}^0$), {\it i.e.}, the sum of the two
individual rates.
We adopt this convention, {\it i.e.}, our FF refer to the fragmentation
of any given parton into $K_S^0$ and $K_L^0$ (or $K^0$ and $\overline K^0$).
For the fitting procedure, we use the $x$~bins in the interval between
$x_{\rm min}=\max(0.1,\,2\,{\rm GeV}/\sqrt{s})$ and $x_{\rm max}=0.8$
and integrate the theoretical functions over the bin widths,
which is equivalent to the experimental binning procedure.
The restriction at small $x$ is to exclude events in the
non-perturbative region, where mass effects are important.
Very-large-$x$ data suffer from huge uncertainties, so we prefer to
disregard the few data-points above $x_{\rm max}$.
As usual, we parameterize the
$x$~dependence of the FF at the starting scale $Q_0$ as
\begin{equation}
\label{ansatz}
D_a^{K^0}(x,Q_0^2)=Nx^{\alpha}(1-x)^{\beta},
\end{equation}
where $a$ stands for any quark flavour or the gluon. We impose the
condition\linebreak
$D_s^{K^0+\overline K^0}(x,Q^2)=D_d^{K^0+\overline K^0}(x,Q^2)$.
For all the other quark~FF, we take $N$, $\alpha$, and $\beta$ to be
independent fit parameters.

As mentioned above, the $e^+e^-$ data on inclusive single particle
production does not well constrain the gluon FF, which, however, plays an
important r\^ole in $ep$ reactions and even more so in $p\bar{p}$ processes.
Since, at present, there exists no additional information on gluon
fragmentation to neutral kaons in $e^+e^-$ annihilation,
we fall back on the results on the fragmentation of gluons into charged kaons
obtained in our recent analysis \cite{bkk2}.
We argue that the supposedly flavour-blind gluon should fragment
into charged and neutral kaons at the same rate, and identify the
corresponding FF.
Later on in this section, we shall demonstrate in more detail
that, in want of better data, this is a sensible prescription.

Of course, the data on $\pi^\pm$ and $K^\pm$ production have much better
statistics than the $K^0$ production data under investigation in this paper.
For this reason, and for compatibility with our $\pi^\pm$ and $K^\pm$ sets,
we do not fit $\Lambda_{\overline{\rm MS}}$ anew, but
adopt the values determined in Ref.~\cite{bkk2},
$\Lambda_{\overline{\rm MS}}^{(5)}=$~107~MeV (195~MeV) in LO (NLO).
We are thus left with 12 independent fit parameters.

The quality of the fit is measured in terms of the $\chi^2_{\rm d.o.f.}$ for
all selected data points. The technical procedure to determine these 12
parameters, using well-tested numerical techniques of multidimensional
optimization \cite{minuit}, is similar to our earlier work \cite{bkk2}.
As in Ref.~\cite{bkk2}, we choose $Q_0=\sqrt{2}$~GeV for the $u$, $d$, and
$s$ quarks, $Q_0=m(\eta_c)=2.9788$~GeV \cite{pdg} for the $c$ quark,
and $Q_0=m(\Upsilon)=9.46037$~GeV \cite{pdg} for the $b$ quark.
Our results are listed below.
For the sum of $K^0$ and $\overline K^0$, we find
\begin{eqnarray}
\label{dlo}
D_u^{(K^0+\overline K^0,\,{\rm LO})}(x,Q_0^2)&\n
=\n&0.51\,x^{-0.841}\,(1-x)^{1.55},\nonumber\\
D_d^{(K^0+\overline K^0,\,{\rm LO})}(x,Q_0^2)&\n
=\n&D_s^{(K^0+\overline K^0,\,{\rm LO})}(x,Q_0^2)
=1.47\,x^{-0.691}\,(1-x)^{3.49},\nonumber\\
D_c^{(K^0+\overline K^0,\,{\rm LO})}(x,Q_0^2)&\n
=\n&1.00\,x^{-0.738}\,(1-x)^{2.93},\nonumber\\
D_b^{(K^0+\overline K^0,\,{\rm LO})}(x,Q_0^2)&\n
=\n&0.68\,x^{-0.598}\,(1-x)^{1.93},\nonumber\\
D_g^{(K^0+\overline K^0,\,{\rm LO})}(x,Q_0^2)&\n
=\n&0.43\,x^{-0.374}\,(1-x)^{2.69}
\end{eqnarray}
in LO and
\begin{eqnarray}
\label{dnlo}
D_u^{(K^0+\overline K^0,\,{\rm NLO})}(x,Q_0^2)&\n
=\n&0.50\,x^{-0.781}\,(1-x)^{1.58},\nonumber\\
D_d^{(K^0+\overline K^0,\,{\rm NLO})}(x,Q_0^2)&\n
=\n&D_s^{(K^0+\overline K^0,\,{\rm NLO})}(x,Q_0^2)
=1.25\,x^{-0.564}\,(1-x)^{3.33},\nonumber\\
D_c^{(K^0+\overline K^0,\,{\rm NLO})}(x,Q_0^2)&\n
=\n&0.99\,x^{-0.601}\,(1-x)^{3.80},\nonumber\\
D_b^{(K^0+\overline K^0,\,{\rm NLO})}(x,Q_0^2)&\n
=\n&0.53\,x^{-0.571}\,(1-x)^{1.98},\nonumber\\
D_g^{(K^0+\overline K^0,\,{\rm NLO})}(x,Q_0^2)&\n
=\n&0.33\,x^{-0.351}\,(1-x)^{0.65}
\end{eqnarray}
in NLO.
Here, it is understood that the $Q_0^2$ values refer to the individual
starting points given above.
For the data that we fitted to, we find very small
$\chi^2_{\rm d.o.f.}$ values, namely 0.53 (0.52) at NLO (LO).
The $\chi^2_{\rm d.o.f.}$ values achieved for the various data sets may
be seen from Table~I.
Our FF also give a good description of the $Z$-resonance data from
DELPHI \cite{delphi} and OPAL \cite{opal}, which yield just slightly larger
values of $\chi^2_{\rm d.o.f.}$
The same is true for the lower-energy data taken by CELLO \cite{cello} and
TASSO \cite{tasso} at PETRA
($\sqrt{s}=35$~GeV) and for the data collected by HRS \cite{hrs} and
TPC \cite{tpc} at PEP ($\sqrt{s}=29$~GeV).
Among the data that we compared with, those from CLEO \cite{cleo} and
ARGUS \cite{argus} have the lowest energy ($\sqrt{s}=10$~GeV).
Only the ARGUS data give an exceptionally large $\chi^2_{\rm d.o.f.}$,
of order 5.

\begin{table}[ht] {TABLE~I. CM energies, experimental collaborations, numbers
of data points used, and $\chi^2_{\rm d.o.f.}$ values obtained at NLO and LO
for the various $e^+e^-$ data samples discussed in the text.
The data used in the fits are marked by an asterisk.}\\[1ex]
\begin{tabular}{|c|lc|c|c|c|c|} \hline
$\sqrt s$ [GeV] & Experiment &&Ref.\ & No.\ of points & $\chi^2_{\rm d.o.f.}$
in NLO & $\chi^2_{\rm d.o.f.}$ in LO \\
\hline
91.2 & ALEPH    &*& \cite{aleph}  &  9 & 0.60 & 0.57 \\
     & DELPHI    && \cite{delphi} & 11 & 0.94 & 0.96 \\
     & OPAL      && \cite{opal}   &  8 & 0.95 & 0.93 \\
\hline
35.0 & CELLO     && \cite{cello}  &  6 & 0.23 & 0.23 \\
     & TASSO     && \cite{tasso}  & 10 & 1.84 & 1.74 \\
\hline
29.0 & MARK~II  &*& \cite{markii} & 11 & 0.48 & 0.49 \\
     & HRS       && \cite{hrs}    & 12 & 2.66 & 2.93 \\
     & TPC       && \cite{tpc}    &  6 & 0.41 & 0.43 \\
\hline
10.49& CLEO      && \cite{cleo}   & 12 & 1.27 & 1.15 \\
\hline
9.98 & ARGUS     && \cite{argus}  &  4 & 5.42 & 5.32 \\
\hline
\end{tabular}
\end{table}

For the reader's convenience, we list simple parameterizations of the $x$ and
$Q^2$ dependences of our $K^0$ FF sets in the Appendix.
We believe that such parameterizations are indispensable for practical
purposes, especially at NLO.
However, we should caution the reader that these parameterizations describe
the evolution of the FF only approximately.
Deviations in excess of 8\% may occur for $x<0.1$ and for $Q>100$~GeV,
in particular for the gluon.
While this kind of accuracy is fully satisfactory for most applications, it is
insufficient for the comparison with the high-statistics data collected at LEP.
We wish to point out that all $\chi^2_{\rm d.o.f.}$ values presented in this
paper have been computed using FF with explicit $Q^2$ evolution, which
have an estimated relative error of less than 0.4\%.

Since we have built in the $c\bar{c}$ and $b\bar{b}$~thresholds, we have three
different starting scales $Q_0$. To illustrate the relative size of the FF for
the different quark flavours and the gluon, we have plotted them in
Fig.~\ref{fig1}
as functions of $x$ for $Q=10$~GeV. We show only the NLO results. The pattern
is somewhat unusual and, contrary to na\"\i ve expectations,
not very similar to the $K^\pm$~FF in our earlier
work \cite{bkk2}. The $u$-quark, $b$-quark, and gluon
FF are rather hard, while the $d/s$-quark and the $c$-quark distributions are
soft. This pattern is already visible at the starting scales $Q_0$ in
Eqs.~(\ref{dlo}) and (\ref{dnlo}), where we must keep in mind, however,
that the starting scale $Q_0$ takes on three distinct values for the light,
$c$, and $b$ quarks. Guided by our findings in connection with $K^\pm$
fragmentation,
we would expect that, in Fig.~\ref{fig1}, the $d/s$-quark FF should be hardest,
and that the $b$-quark FF should resemble that of the $c$ quark.
At this stage, it cannot be excluded that the relative importance of the
individual quark flavours will need some adjustment.
But for this we would need additional $e^+e^-$~data on inclusive $K^0$
production for which the fragmentation of the various quark flavours and the
gluon is disentangled, similarly to what has been done in the case
of charged-hadron production.
Unfortunately, the existing information from $ep$ and $p\bar{p}$ collisions
does not help us much either. Due to its high threshold, $b$-quark production
is absent at small $p_T$, below 9.5~GeV.
In our $ep$ analysis, $c/\bar{c}$ production accounts for 8\% (10\%) of the
cross section at $p_T=5$ (8)~GeV, while, in our $p\bar p$ calculation for
$\sqrt s=1.8$~TeV, its contribution at the same $p_T$ values
is 0.7\% (0.8\%), {\it i.e.}, in both reactions it is small or negligible.

The goodness of our fits to the ALEPH \cite{aleph} and MARK~II \cite{markii}
data may be judged from Fig.~\ref{fig2}. At NLO (LO), we find
$\chi^2_{\rm d.o.f.}$ values of 0.60 (0.57) for ALEPH and 0.48 (0.49) for
MARK~II.

The factorization theorem guarantees that the FF which
we extracted from $e^+e^-$ data may also be used to predict other types of
inclusive single $K^0$ production cross sections, {\it e.g.}, for
$\gamma\gamma$, $ep$, or hadron-hadron collisions. In the following, we shall
present NLO predictions for inclusive photoproduction of $K_S^0$ mesons at
HERA and confront them with preliminary data taken by H1 \cite{linsel}.
As in the H1 measurement, we shall consider the $p_T$~spectrum of the produced
$K_S^0$ mesons, averaged over the rapidity range $|y_{\rm lab}|<1.5$.
We shall work at NLO in the $\overline{\rm MS}$~scheme with $N_f=5$ quark
flavours, fix the renormalization and factorization scales by setting
$\mu=M_{\gamma}=M_p=M_h=\xi p_T$, and adopt the NLO parton distribution
functions (PDF) of the photon and the proton from Refs.~\cite{grv} and
\cite{lai}, respectively, together with our NLO FF.
We wish to emphasize that also the hard-scattering cross sections will be
calculated up to NLO.
We shall evaluate $\alpha_s$ to two loops
with $\Lambda_{\overline{\rm MS}}^{(5)}=158$~MeV \cite{lai}.
The quasi-real photon spectrum will be simulated according to H1 conditions,
by imposing the cut $0.3<z<0.7$ on $z=E_{\gamma}/E_e$ and choosing
$Q^2_{\rm max}=0.01$~GeV$^2$. Our predictions
for $\xi=1/2$, 1, and $2$ are
confronted with the H1 points in Fig.~\ref{fig3}.
The agreement is satisfactory as for both shape and normalization.
Unfortunately, the H1 data are accumulated at rather small $p_T$
($p_T \le 3$~GeV),
whereas our predictions should be more reliable at larger $p_T$. We must keep
in mind, however, that this represents the first measurement of inclusive
$K_S^0$ production at HERA, based on data taken in 1993, and that the numbers
are still preliminary.
More data at larger $p_T$ is expected to appear after
the analysis of the 1994 run is completed. As we see in Fig.~\ref{fig3}, the
cross section shows only moderate scale dependence, which indicates relatively
good perturbative stability. Notice that our prediction in Fig.~\ref{fig3}
refers to $K_S^0$~production, which corresponds to the average of $K^0$ and
$\overline K^0$.

There only exists rather limited experimental information on inclusive
$K_S^0$~production
in $p\bar{p}$~collisions. The only high-energy data available come from the UA5
Collaboration \cite{ua5} at the CERN S$p\bar{p}$S Collider and from the CDF
Collaboration \cite{cdf} at the Fermilab Tevatron.
In Fig.~\ref{fig4}, we show our predictions for the $p_T$ spectrum of
$p\bar{p}\to K_S^0+X$ at $\sqrt{s}=200$, 546, and 900~GeV, with
rapidity averaged over the interval $-2.5<y<2.5$.
The calculation is performed at NLO in the $\overline{\rm MS}$ scheme
with $N_f=5$ quark flavours using the CTEQ3 proton PDF \cite{lai}.
The renormalization and fragmentation scales are identified and set equal
to $p_T/2$, $p_T$, and $2p_T$. The agreement with the UA5 data \cite{ua5} is
quite satisfactory. It is best for the highest CM energy.
Unfortunately, this data is accumulated at rather small $p_T$.
On average, the agreement is
best with scales equal to $p_T$. The data from CDF \cite{cdf} is more recent.
It was taken at $\sqrt{s}=630$ and 1800~GeV. This data together with our
theoretical results for scales $p_T/2$, $p_T$, and $2p_T$
are plotted versus $p_T$ in Fig.~\ref{fig5}. The experimental and theoretical
results are both averaged over $|y|<1.0$. Unfortunately, the CDF data has
rather small $p_T$, too. Again, the
agreement of our calculation with the data is best for scales equal to $p_T$.
Only at very small $p_T$, deviations
from the data taken at $\sqrt{s}=1800$~GeV are noticeable. This is to be
expected, since the theoretical predictions are valid only for
large $p_T$; their reliability at $p_T$ below 1.5~GeV is certainly
questionable.
At this point, we would like to encourage our experimental colleagues in the
CDF Collaboration to also analyze the vast amount of data collected after 1989
with respect to light-meson fragmentation.
In view of the considerable recent theoretical progress in this field,
this would be interesting and exciting in its own right,
rather than but a boring measure to assess backgrounds for certain other
processes which presently happen to be more en vogue.
In fact, this would allow us to test the QCD-improved parton model and, in
particular, the factorization theorem at the quantum level.

Due to their limited $p_T$ range and their modest accuracy,
the data sets presented in Figs.~\ref{fig4} and \ref{fig5} are not so well
suited for constraining the FF obtained from the $e^+e^-$~analysis.
However, they provide a welcome cross check, in particular with respect to
the gluon FF, which is only feebly constrained by the $e^+e^-$ data.
To elaborate this point, we investigate the influence of the gluon
fragmentation on the $ep$ and $p\bar{p}$ cross sections.
To that end, we repeat the calculations of Figs.~\ref{fig3}--\ref{fig5}
switching off the quark FF.
In Fig.~\ref{fig6}, we show the outcome normalized to the full calculations
for the $ep$ cross section and the 200~GeV, 630~GeV, and 1800~GeV
$p\bar{p}$~cross sections.
We observe that, in the low-$p_T$ range, the $p\bar{p}$ cross sections are
overwhelmingly dominated by the gluon FF.
The ratio increases with CM energy and reaches 90\% at the largest energy.
This shows that, if it were not for the large errors, the $p\bar{p}$ data
would be perfectly well suited for constraining the gluon FF.
Looking back at Figs.~\ref{fig4} and \ref{fig5}, it is fair to say that
the strength of the gluon FF as obtained from our
$e^+e^-\to h^{\pm} + X$ fits is large enough to account for the
$p\bar{p}$ data.
This is in accord with recent studies of inclusive
charged-hadron production in $p\bar{p}$ collisions \cite{bk}. We also
examined in which $x$~range the gluon FF maximally contributes to the
$p\bar{p}$ inclusive cross sections in the considered $p_T$ range.
Depending on the CM energy, the most important $x$ values are concentrated
around $x=0.4$. This means that the
$p\bar{p}$ data only constrains the gluon FF in a limited range of $x$. On the
other hand, we know that the $e^+e^-$ data does not determine the gluon FF very
accurately, {\it i.e.}, a good description of the $e^+e^-$ data may also be
obtained with a weaker gluon FF.
The $ep$ data also need a sufficiently strong gluon FF, in
particular to describe the data near $p_T=2$~GeV. At larger $p_T$, the
influence of the gluon FF diminishes, and the quark FF come into play much
more strongly. This is to be expected, since, in the $ep$ cross section,
the $qg\to qg$ channel is similarly important as the $gg\to gg$ and $qg\to gq$
channels, even at small $p_T$.

Having established the importance of the gluon FF for $K_S^0$ production
in $p\bar{p}$ collisions, we should take a closer look at our assumptions
concerning the gluon FF. These were twofold. We explicitly stated that we
were going to assume $D_g^{K^0}(x,Q_0)=D_g^{K^{\pm}}(x,Q_0)$ (a). A second,
hidden assumption was that $D_g^{K^{\pm}}$ had been well constrained by our
previous analysis \cite{bkk2}, although only experimental information on the
gluon FF for the sum of the charged hadrons had been available (b).
By investigating the ratio of the cross section for inclusive kaon production
in hadron collisions to that for charged hadrons,
we may check both assumptions. For one thing,
this ratio is approximately equal to the ratio of the respective gluon FF
(at least for high CM energies), which enables us to test (b).
On the other hand, this ratio has been measured for both charged and
neutral kaons, providing us with a check of (a).
In Fig.~\ref{fig7}, we confront our predictions, based on assumptions (a) and
(b), with the experimental data on neutral-kaon production in 1.8~TeV $p\bar p$
collisions by CDF \cite{cdf} and on charged-kaon production in 53~GeV and
27~GeV $pp$ scattering by the British--Scandinavian Collaboration \cite{bs} at
the CERN ISR and by the Chicago--Princeton Collaboration \cite{cp} at
Fermilab, respectively.
In the theoretical calculation of charged-hadron production,
only charged pions and kaons are included.
Protons, $\Lambda$ hyperons, and other heavy hadrons
are known to contribute little to the cross section and are
neglected here. We find throughout good agreement with the data.
All data, as well as our predictions, approach a plateau
at not-too-small $p_T$. Its height is about 0.2,
fairly independently of the CM energy or whether neutral or
charged kaons are considered.

At this point, we should compare our results on the $K^0$ FF with those
obtained in Ref.~\cite{gre}.
The cross section of inclusive $K_S^0$ production at $\sqrt s=1.8$~TeV
under CDF conditions as predicted by Ref.~\cite{gre} agrees reasonably
well with our own calculation; it ranges between 50\% and 70\% of our result
for $p_T$ between 2 and 7~GeV.

These tests reassure us of the soundness of the assumptions concerning the
gluon FF of the neutral kaons which we had to make in view of shortcomings in
the presently available experimental information.
{}From a theoretical point of view, it would certainly be desirable to
constrain
the $K^0$ FF by using just $e^+e^-$ data, as this would enable us to test them
in other types of processes so as to probe the factorization theorem.
Unfortunately, this is not yet possible, which has led us to use
additional input to obtain FF that satisfactorily describe a variety of
$e^+e^-$, $ep$, and $p\bar{p}$ data.

\section{Summary and Conclusions}

We presented FF for neutral kaons, both at LO and NLO. They were constructed
from fits to data on inclusive $K^0+\overline{K}^0$ production in $e^+e^-$
annihilation taken by MARK~II \cite{markii} at PEP ($\sqrt{s}=29$~GeV) and by
ALEPH at LEP \cite{aleph}. Although our FF were only fitted to the MARK~II and
ALEPH data, it turned out that they lead to an excellent description of other
$e^+e^-$ data on inclusive $K^0+\overline{K}^0$ production ranging from
$\sqrt{s}=10$~GeV to LEP energy. We always obtained $\chi^2_{\rm d.o.f.}$
values of order unity. The only exception, with
$\chi^2_{\rm d.o.f.} \approx 5$, occurred
for the ARGUS data at $\sqrt{s}=9.98$~GeV.

Since the $e^+e^-$ data do not constrain the gluon FF so well,
we made NLO predictions for the $p_T$ spectra of $K_S^0$ mesons produced
inclusively in the scattering of quasi-real photons on protons under HERA
conditions and in proton-antiproton collisions under UA5 and CDF conditions,
and confronted them with the respective data. The agreement
turned out to be quite satisfactory. We discovered that the gluon FF is very
important to account for the $p\bar{p}$ data. We are thus faced with the
unfortunate situation that the $p\bar{p}$ data almost exclusively tests the
gluon FF, which is of little relevance for existing $e^+e^-$ data.
Vice versa, the quark FF, which---up to a residual uncertainty in the
relative importance of the individual flavours---are fixed
by a wealth of $e^+e^-$ data, have hardly any impact on
$p\bar{p}\to K_S^0 + X$.
The situation will be ameliorated as soon as the $ep$ scattering experiments at
HERA provide us with higher-statistics data, in particular at larger $p_T$.
In conclusion, present data does not yet allow us
to test the universality of the FF in inclusive $K^0$ production;
the situation rather requires that we exploit the universality postulated by
the factorization theorem in order to extract meaningful FF.
This was achieved in the work presented here.

In order to make further progress, one would need $e^+e^-$ data on inclusive
$K^0$ production in which the different quark flavours are tagged.
Also, the gluon FF would have to be constrained better, {\it e.g.}, by
studying inclusive $K^0$ production in tagged three-jet events or by measuring
the longitudinal part of the cross section, similarly to what has been
done for charged particles.
As for $ep$ and $p\bar{p}$ collisions, data at larger $p_T$ with sufficient
accuracy would be highly welcome, since they would allow us to quantitatively
test the factorization theorem of fragmentation in the
QCD-improved parton model.

\bigskip
\centerline{\bf ACKNOWLEDGMENTS}
\smallskip\noindent

We are grateful to Frank Linsel for making available to us the preliminary
H1 data on inclusive photoproduction of neutral kaons prior to their
official publication \cite{linsel}.
We thank Simona Rolli for providing us with numerical results obtained on
the basis of Ref.~\cite{gre} and for useful discussions.
One of us (BAK) is indebted to the FNAL Theory Group for inviting him as a
Guest Scientist and for the great hospitality extended to him.

\begin{appendix}

\section{Parameterizations}

For the reader's convenience, we shall present here simple parameterizations
of the $x$ and $Q^2$ dependence of our FF.\footnote{%
A FORTRAN subroutine that returns the FF for given $x$ and $Q^2$ may be
obtained from the authors via e-mail (binnewie@ips107.desy.de,
kniehl@vms.mppmu.mpg.de).}
As usual, we introduce the scaling variable
\begin{equation}
\label{sbar}
\bar{s}=\ln{\frac{\ln{(Q^2/\Lambda^2)}}{\ln{(Q^2_0/\Lambda^2)}}}.
\end{equation}
For $\Lambda$ we use the $\overline{\rm MS}$ value appropriate to $N_f=5$
flavours, since the parameterization would not benefit from the incorporation
of discontinuities in $\bar{s}$.
$\Lambda_{\overline{\rm MS}}^{(5)}$ is taken from our previous fit
\cite{bkk2} to be
107~MeV (195~MeV) in LO (NLO).
Similarly to Eqs.~(\ref{dlo})--(\ref{dnlo}), we use three different values
for $Q_0$, namely
\begin{eqnarray}
Q_0&=&\left\{\begin{array}{l@{\quad\mbox{if}\quad}l}
\sqrt2~\mbox{GeV}, & a=u,d,s,g\\
m(\eta_c)=2.9788~\mbox{GeV}, & a=c\\
m(\Upsilon)=9.46037~\mbox{GeV}, & a=b
\end{array}\right.\;.
\end{eqnarray}
This leads to three different definitions of $\bar s$.
For definiteness, we use the symbol $\bar s_c$ for charm and $\bar s_b$ for
bottom along with $\bar s$ for the residual partons.

We parameterize our FF by simple functions in $x$ with coefficients
which we write as polynomials in $\bar s$, $\bar s_c$, and $\bar s_b$.
We find that the template
\begin{equation}
\label{temp}
D(x,Q^2)=Nx^{\alpha}(1-x)^{\beta}
\end{equation}
is sufficiently flexible,
except for $D_g^{(K^0+{\overline K}^0,\,{\rm NLO})}$,
where we include an additional factor $(1+\gamma/x)$ on the right-hand side of
Eq.~(\ref{temp}).
For $\bar s=\bar s_c=\bar s_b=0$, the parameterizations agree with
the respective ans\"atze in Eqs.~(\ref{dlo})--(\ref{dnlo}).
The charm and bottom parameterizations must be put to zero by hand for
$\bar s_c<0$ and $\bar s_b<0$, respectively.

We list below the parameters to be inserted in Eq.~(\ref{temp})
both at LO and NLO.
The resulting parameterizations correctly describe the $Q^2$ evolution up to
8\% for $Q_0\le Q\le100$~GeV and $0.1\le x\le0.8$.
\begin{enumerate}
\item LO FF for $(K^0+{\overline K}^0)$:
\begin{itemize}
\item $D_u^{(K^0+{\overline K}^0,\,{\rm LO})}(x,Q^2)$:
\begin{eqnarray}
N&\n=\n& 0.510 -0.251\bar s +0.036\bar s^2 \nonumber\\
\alpha&\n=\n& -0.841 -0.285\bar s +0.021\bar s^2 \nonumber\\
\beta&\n=\n&1.550 +0.712\bar s -0.069\bar s^2 +0.037\bar s^3
\end{eqnarray}
\item $D_d^{(K^0+{\overline K}^0,\,{\rm LO})}(x,Q^2)
=D_s^{(K^0+{\overline K}^0,\,{\rm LO})}(x,Q^2)$:
\begin{eqnarray}
N&\n=\n&1.470-1.088\bar s +0.276\bar s^2 \nonumber\\
\alpha&\n=\n& -0.691 -0.312\bar s -0.045\bar s^2 +0.038\bar s^3\nonumber\\
\beta&\n=\n&3.490 +0.851\bar s -0.181\bar s^2 +0.098\bar s^3
\end{eqnarray}
\item $D_c^{(K^0+{\overline K}^0,\,{\rm LO})}(x,Q^2)$:
\begin{eqnarray}
N&\n=\n&1.000 -0.679\bar s_c +0.111\bar s_c^2 +0.058\bar s_c^3\nonumber\\
\alpha&\n=\n& -0.738 -0.302\bar s_c -0.073\bar s_c^2 +0.084\bar
s_c^3\nonumber\\
\beta&\n=\n&2.930 +0.758\bar s_c -0.118\bar s_c^2 +0.107\bar s_c^3
\end{eqnarray}
\item $D_b^{(K^0+{\overline K}^0,\,{\rm LO})}(x,Q^2)$:
\begin{eqnarray}
N&\n=\n& 0.680 -0.470\bar s_b +0.241\bar s_b^2 -0.115\bar s_b^3\nonumber\\
\alpha&\n=\n& -0.598 -0.407\bar s_b +0.059\bar s_b^2 +0.063\bar
s_b^3\nonumber\\
\beta&\n=\n&1.930 +0.554\bar s_b +0.232\bar s_b^2 -0.186\bar s_b^3
\end{eqnarray}
\item $D_g^{(K^0+{\overline K}^0,\,{\rm LO})}(x,Q^2)$:
\begin{eqnarray}
N&\n=\n& 0.430 -0.881\bar s +0.860\bar s^2 -0.320\bar s^3\nonumber\\
\alpha&\n=\n& -0.374 -2.147\bar s+1.239\bar s^2 -0.232\bar s^3\nonumber\\
\beta&\n=\n&2.690+1.515\bar s -0.188\bar s^2 -0.071\bar s^3
\end{eqnarray}
\end{itemize}
\item NLO FF for $(K^0+{\overline K}^0)$:
\begin{itemize}
\item $D_u^{(K^0+{\overline K}^0,\,{\rm NLO})}(x,Q^2)$:
\begin{eqnarray}
N&\n=\n& 0.500 -0.125\bar s -0.051\bar s^2 \nonumber\\
\alpha&\n=\n& -0.781 -0.500\bar s +0.264\bar s^2 -0.133\bar s^3\nonumber\\
\beta&\n=\n&1.580+1.074\bar s -0.380\bar s^2 +0.084\bar s^3
\end{eqnarray}
\item $D_d^{(K^0+{\overline K}^0,\,{\rm NLO})}(x,Q^2)
=D_s^{(K^0+{\overline K}^0,\,{\rm NLO})}(x,Q^2)$:
\begin{eqnarray}
N&\n=\n&1.250-1.385\bar s +0.896\bar s^2 -0.251\bar s^3\nonumber\\
\alpha&\n=\n& -0.564 -0.718\bar s +0.208\bar s^2 \nonumber\\
\beta&\n=\n&3.330 +0.100\bar s +0.388\bar s^2 -0.064\bar s^3
\end{eqnarray}
\item $D_c^{(K^0+{\overline K}^0,\,{\rm NLO})}(x,Q^2)$:
\begin{eqnarray}
N&\n=\n& 0.990-1.020\bar s_c +0.639\bar s_c^2 -0.199\bar s_c^3\nonumber\\
\alpha&\n=\n& -0.601 -0.666\bar s_c +0.189\bar s_c^2 \nonumber\\
\beta&\n=\n&3.800 +0.192\bar s_c +0.382\bar s_c^2 -0.077\bar s_c^3
\end{eqnarray}
\item $D_b^{(K^0+{\overline K}^0,\,{\rm NLO})}(x,Q^2)$:
\begin{eqnarray}
N&\n=\n& 0.530 -0.541\bar s_b +0.429\bar s_b^2 -0.171\bar s_b^3\nonumber\\
\alpha&\n=\n& -0.571 -0.803\bar s_b +0.371\bar s_b^2 \nonumber\\
\beta&\n=\n&1.980 +0.308\bar s_b +0.298\bar s_b^2 -0.035\bar s_b^3
\end{eqnarray}
\item $D_g^{(K^0+{\overline K}^0,\,{\rm NLO})}(x,Q^2)$:
\begin{eqnarray}
N&\n=\n& 0.330 -0.307\bar s +0.075\bar s^2 \nonumber\\
\alpha&\n=\n& -0.351 -0.040\bar s -0.282\bar s^2 +0.033\bar s^3\nonumber\\
\beta&\n=\n& 0.650+2.141\bar s -0.457\bar s^2 +0.075\bar s^3\nonumber\\
\gamma&\n=\n&  1.096\bar s -0.198\bar s^2
\end{eqnarray}
\end{itemize}
\end{enumerate}
\end{appendix}

\newpage

\vskip-6cm

\begin{figure}

\centerline{\bf FIGURE CAPTIONS}

\caption{\protect\label{fig1} $x$ dependence of the NLO set of
$K^0 + \overline K^0$ FF at $Q^2=100$~GeV$^2$.\hskip5cm}

\vskip-.2cm

\caption{\protect\label{fig2} Differential cross sections of inclusive
$K^0 + \overline K^0$ production at LO (dashed lines) and NLO (solid lines) as
functions of $x$ at $\protect\sqrt{s}=91.2$ and 29.0~GeV.
The theoretical calculations are compared with the respective experimental
data by ALEPH \protect\cite{aleph} and MARK~II \protect\cite{markii}.
For better separation, the distributions at $29.0$~GeV have been divided by
$10$.
\hskip5cm}

\vskip-.2cm

\caption{\protect\label{fig3} The (preliminary) $p_T$ spectrum of
inclusive $K_S^0$ production in $ep$ collisions as measured
by H1 \protect\cite{linsel} is compared with the NLO
calculation in the $\overline{\rm MS}$ scheme
with $N_f=5$ flavours using the photon and proton PDF of
Refs.~\protect\cite{grv} and \protect\cite{lai}, respectively,
together with our FF.
The dashed/solid/dash-dotted curves correspond
to the choices $\xi=0.5/1/2$.
\hskip5cm}

\vskip-.2cm

\caption{\protect\label{fig4}
The $p_T$ spectra of inclusive $K_S^0$ production in $p\bar{p}$
collisions as measured by UA5 \protect\cite{ua5}
at $\protect\sqrt s=900$, 546, and 200~GeV are compared with the respective
NLO calculations in the $\overline{\rm MS}$ scheme
with $N_f=5$ flavours using the proton and antiproton PDF of
Ref.~\protect\cite{lai}.
For better separation, the spectra have been separated by factors of 10.
The dashed/solid/dash-dotted curves correspond
to the choices $\xi=0.5/1/2$.
\hskip5cm}

\vskip-.2cm

\caption{\protect\label{fig5}
Same as Fig.~\protect\ref{fig4}, but for data from CDF \protect\cite{cdf} at
$\protect\sqrt s=1800$ and $630$~GeV.
\hskip5cm}

\vskip-.2cm

\caption{\protect\label{fig6}
Percentage of events with gluon fragmentation in the $p_T$ spectra of
$K_S^0$ mesons inclusively produced in $p\bar p$ and $ep$ collisions.
The solid, dash-dotted, dashed, and dotted lines represent our predictions for
the 1800~GeV and 630~GeV CDF \protect\cite{cdf}, 200~GeV UA5
\protect\cite{ua5}, and H1 \protect\cite{linsel} experiments, respectively.
\hskip5cm}

\vskip-.2cm

\caption{\protect\label{fig7}
Ratio of the differential cross section for inclusive kaon production to that
for charged hadrons as a function of $p_T$.
We compare the $p\bar p$ data on $K_S^0$ mesons by CDF \protect\cite{cdf}
and the $pp$ data on $K^+$ and $K^-$ mesons (averaged) by the
British-Scandinavian (BS) \protect\cite{bs} and Chicago-Princeton (CP)
\protect\cite{cp} Collaborations with the respective NLO calculations using
our FF.
We compute the denominators by summing over the charged pions and kaons.
\hskip8cm}

\end{figure}

\vfill

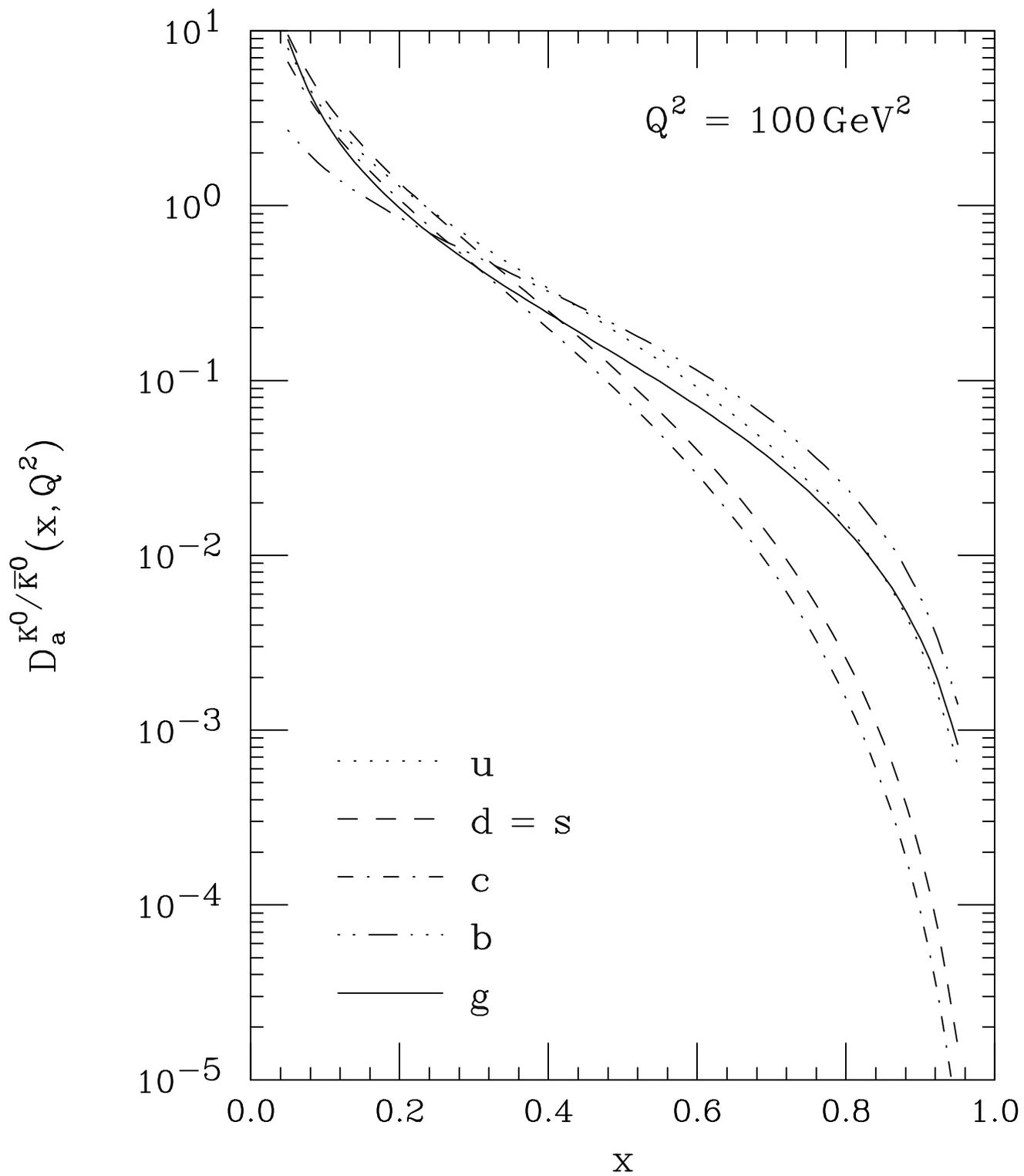
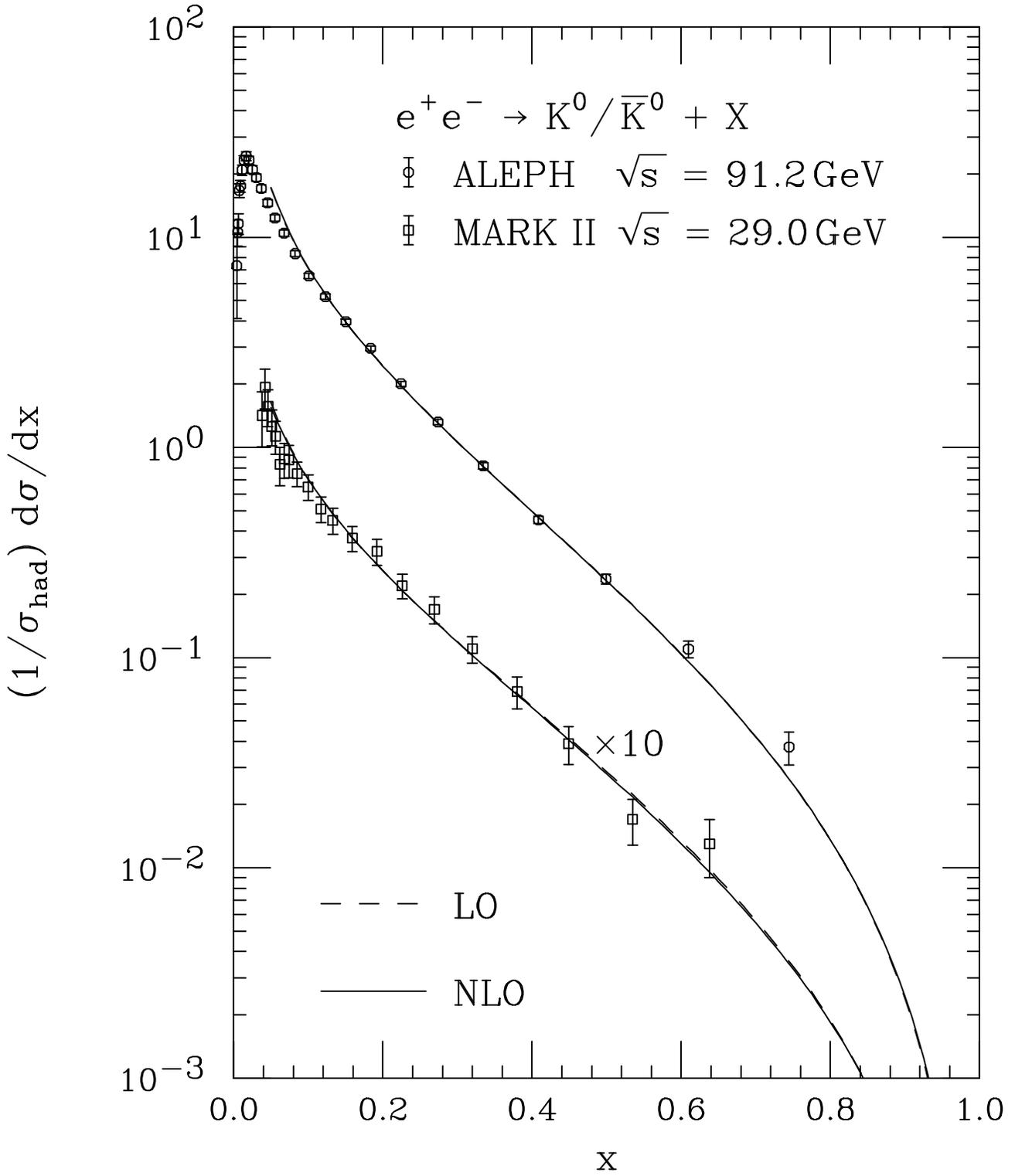
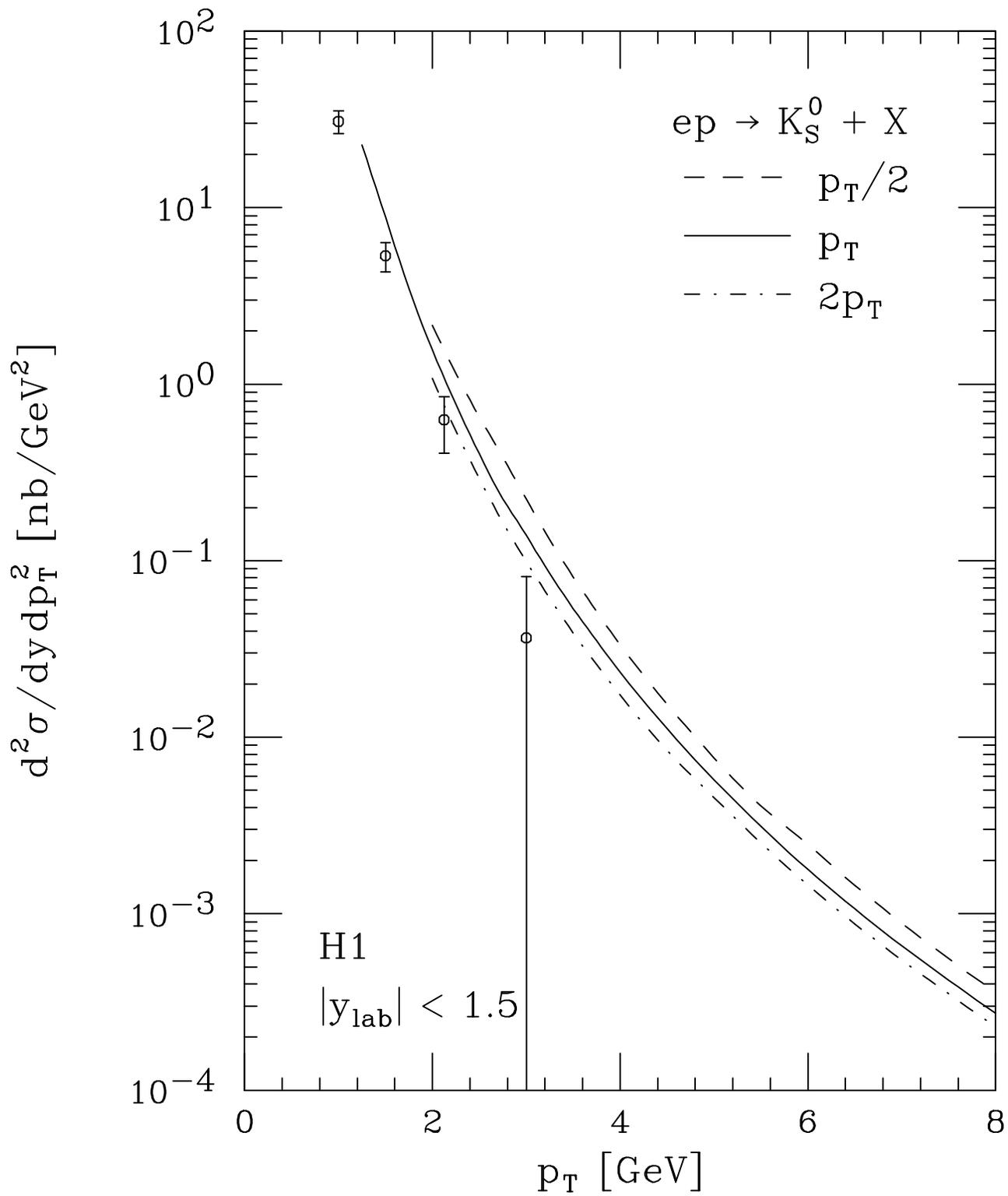
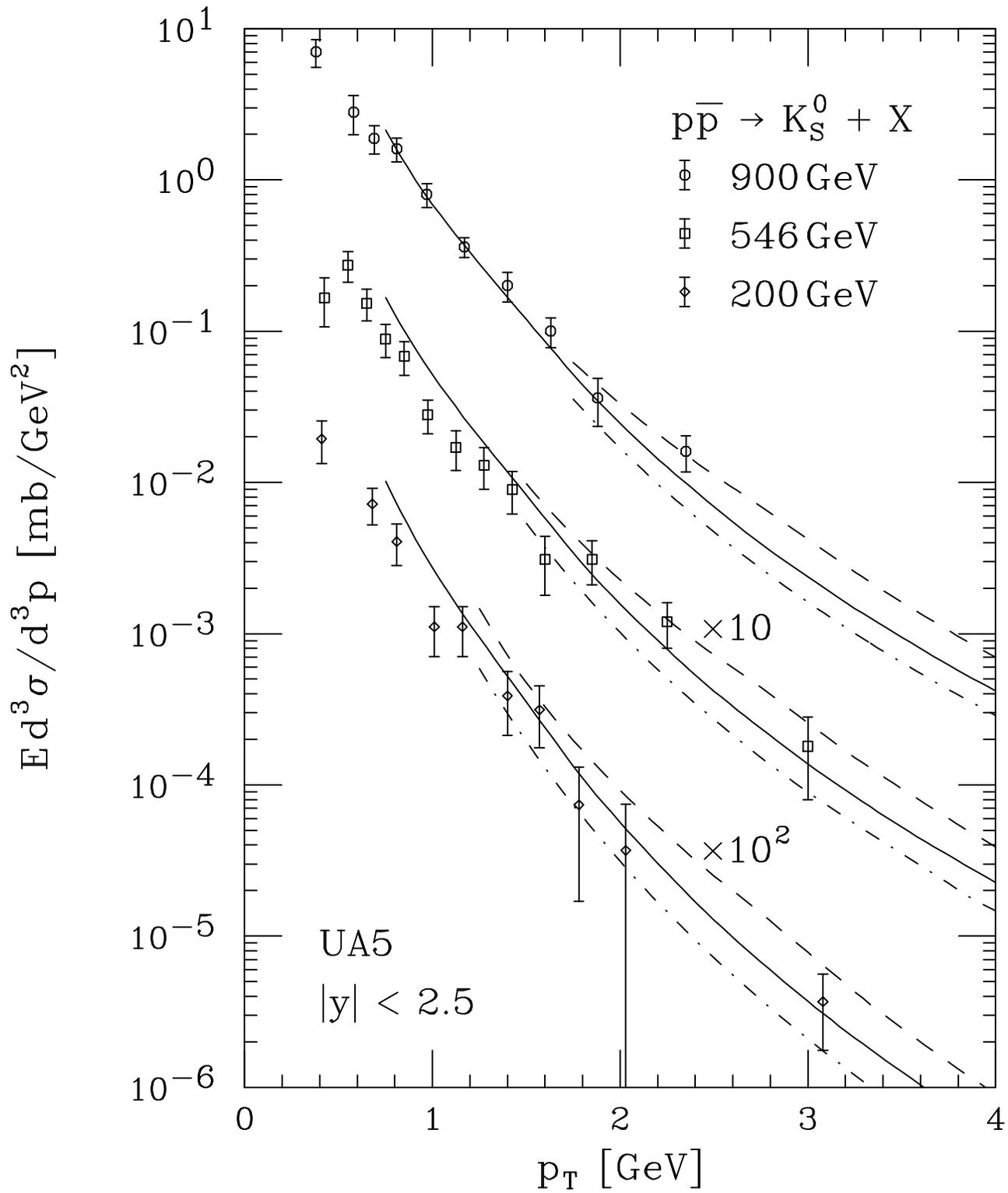
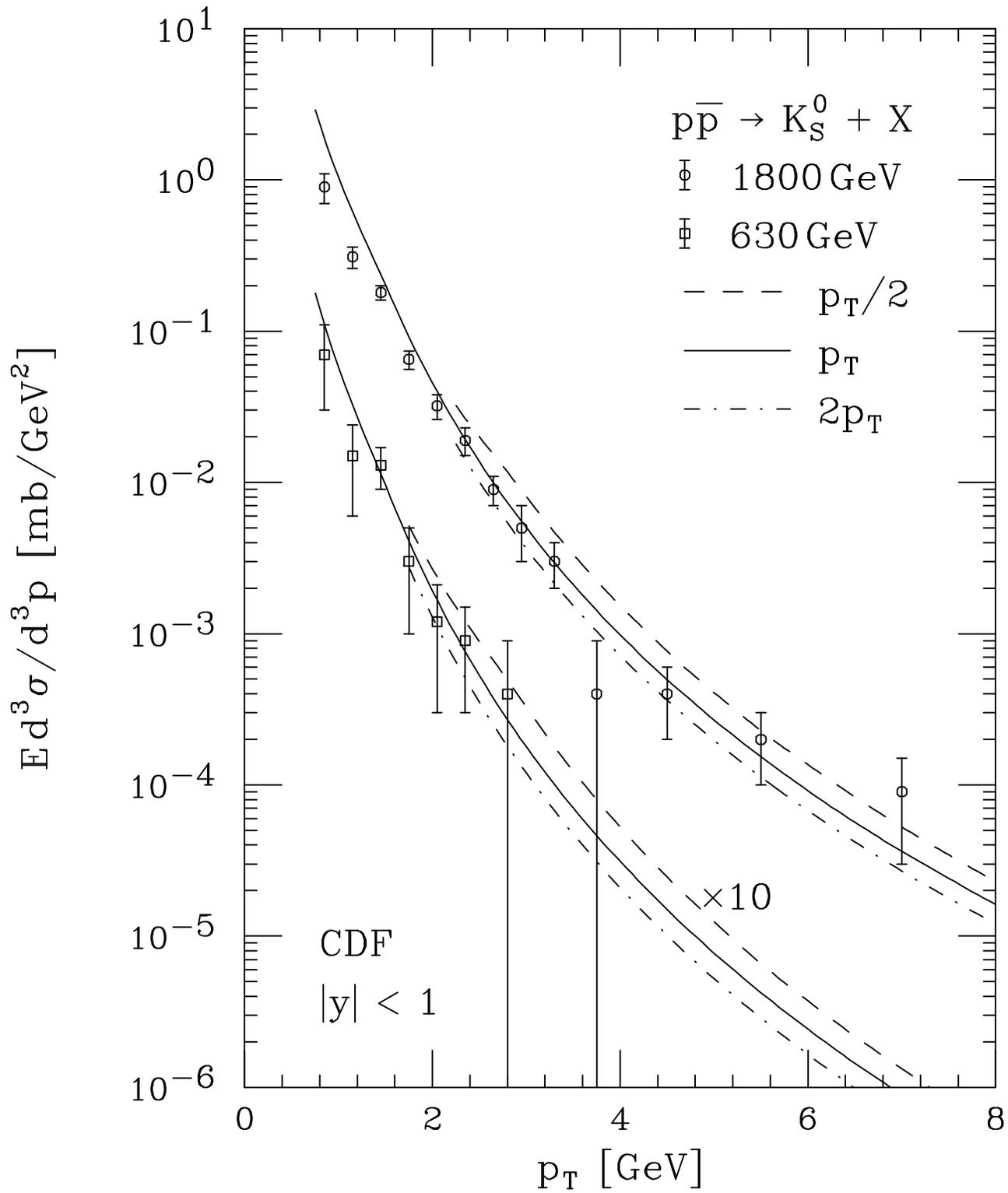
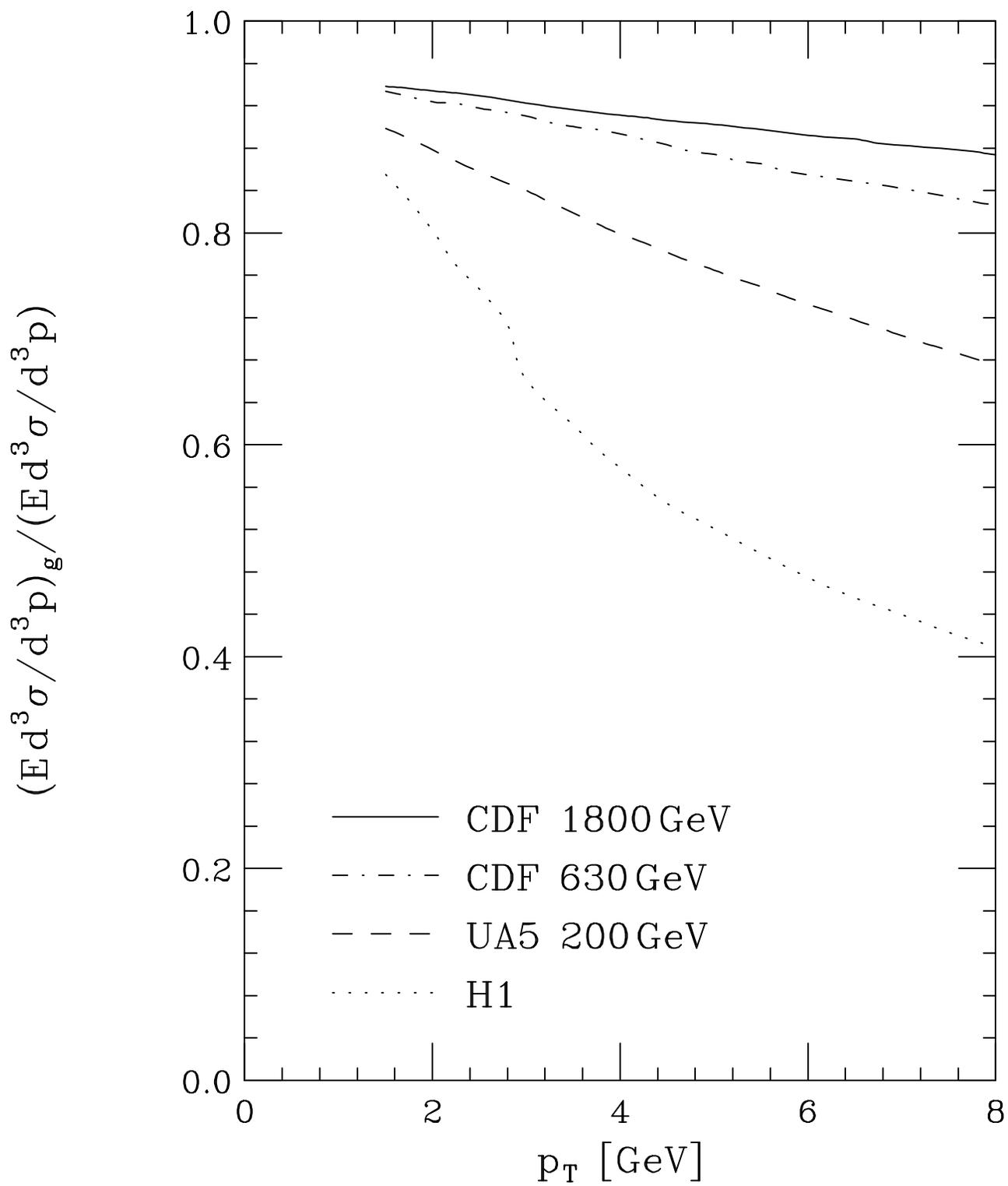
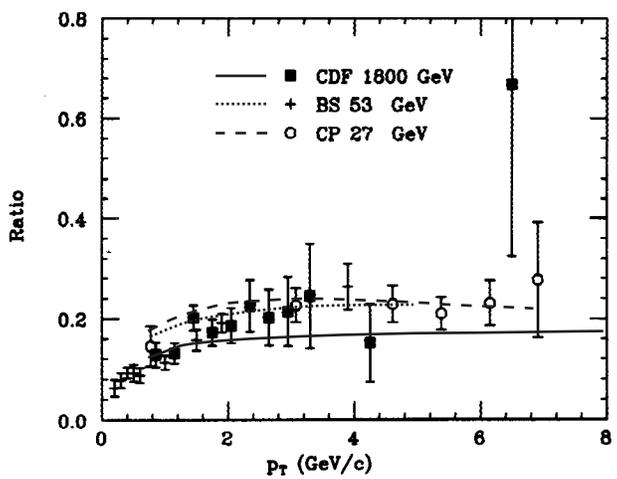
\end{document}